\documentclass[english,aps,prl,amsmath,amssymb,twocolumn]{revtex4}
\usepackage{epsfig}
\usepackage{bm}
\usepackage{bbold}
\usepackage{latexsym}
\usepackage{color}
\usepackage{hyperref}

\begin{document}
	\title{On momentum operators given by Killing vectors \\whose integral curves are geodesics}
	\author{Thomas Sch\"urmann}
	\email{<t.schurmann@icloud.com>}
	\affiliation{J\"ulich Supercomputing Centre, J\"ulich Research Centre, D-52425 J\"ulich, Germany}
\begin{abstract}
The paper considers momentum operators on intrinsically curved manifolds. Given that the momentum operators are Killing vector fields whose integral curves are geodesics, it is shown that the corresponding manifold is either flat, or otherwise of compact type with positive constant sectional curvature and dimension equal to 1, 3 or 7. Explicit representations of momentum operators and the associated Casimir element will be discussed for the 3-sphere $S^3$. It will be verified that the structural constants of the underlying Lie algebra are proportional to $2\hbar/R$, where $R$ is the curvature radius of $S^3$, and $\hbar$ is the reduced Planck's constant. This results in a countable energy and momentum spectrum of freely moving particles in $S^3$. It is shown that the maximum resolution of the possible momenta is given by the de Broglie wave length, $\lambda_R=\pi R$, which is identical to the diameter of the manifold. The corresponding covariant position operators are defined in terms of geodesic normal coordinates and the associated commutator relations of position and momentum are established.
\end{abstract}

\maketitle

\section{I.\, Introduction}
Every generalization of the ordinary momentum operator in quantum mechanics to intrinsically curved manifolds strongly depends on the assumptions which are supposed to be established. Those assumptions are mostly based on the rules of quantum mechanics in Cartesian coordinates of the flat Euclidean space. 

Let $M$ be an $n$-dimensional smooth Riemannian manifold with metric $g$ (occasionally denoted by $\langle \cdot,\cdot\rangle$). At every point $p\in M$, smooth manifolds admit a tangent space, $T_pM$, which is an $n$-dimensional real vector space. For every smooth function $f$ on $M$, consider the action of the differential form, $df$. Since $df$ is a map on the tangential space $T_pM$ at point $p\in M$, the gradient of $f$ is defined such that  
\begin{eqnarray}\label{df}
	df(v)=\langle v,\text{grad} f\rangle
\end{eqnarray}
for every smooth vector field $v\in T_pM$.
From this definition one can obtain the formal expression of the gradient vector field by 
\begin{eqnarray}\label{grad}
	\text{grad} f=(\text{grad} f)^k \,\partial_k
\end{eqnarray}
with contravariant components $(\text{grad} f)^i=g^{ik}\partial_k f$, where $g^{ik}$ are components of the inverse metric $g^{-1}$, $\{\partial_k\}$ is the natural basis of $T_pM$, and the Latin letters denote the space indices, $i,k=1,2,\cdots,n$.
With this notation the traditional momentum operator of Euclidean space in Cartesian coordinates is given by the covariant components $\hat{p}_k f=-i \hbar\,(\text{grad} f)_k$, which is commonly written as 
\begin{eqnarray}\label{pk}	
\hat{p}_k=-i \hbar\,\partial_k
\end{eqnarray}
in physics textbooks; here, $\hbar$ is the reduced Planck's constant.  
Following De Witt \cite{dW52}\cite{dW57}, for intrinsically curved manifolds one can obtain more general momentum operators which also satisfy the canonical commutation relations 
\begin{eqnarray}	
\left[ \hat{x}^j,\hat{p}_k\right] &=& i\hbar\,\delta^j_k\label{c0}\\
\left[\, \hat{p}_j\,, \hat{p}_k \right] &=& \left[\, \hat{x}_j\,, \hat{x}_k \right] = 0.\label{c1}
\end{eqnarray}
This yields an enhanced  quantization rule 
\begin{eqnarray}\label{pdw}	
	\hat{p}_k=-i \hbar \left(\partial_k + \Gamma^j_{jk}(x)\right)
\end{eqnarray}
where the curvature of the manifold is reflected by the contracted Christoffel symbols $\Gamma^j_{jk}(x)$. Compared to the Cartesian quantization rule in Euclidean space (\ref{pk}), the ordinary partial derivative is replaced by the definition (\ref{pdw}). An advantage of this quantization rule is its applicability to a wide range of curved spaces and the validity of the canonical commutator relations. However, in order to set up a quantum Hamiltonian one has to keep in mind that there is no unique prescription to quantize the classical curved space Hamilton function. This is because of operator orderings of the kinetic energy term result in different, inequivalent quantum corrections, such that the correct Hamiltonian can only be confirmed empirically.  
   
On the other hand, a particular property which is in a sense self-evident for the Cartesian case is that the partial derivatives $ \partial_k$ of the momentum representation in (\ref{pk}) can be understood as {\it orthonormal} Killing vectors whose isometries are "translations" in Euclidean space. Strictly speaking, the integral curves of this Killing vectors (called Killing trajectories) are geodesics, especially for a particle moving in a force-free surrounding. If this idea is generalized, for instance, to a freely moving particle on the 3-sphere of radius $R$, then it is already know that the (geodesic) Killing trajectories are the greater circles. Hence, the structural coefficients of the underlying Lie algebra are proportional to the fraction $\hbar/R$ and therefore different from zero and one is compelled to relax the form of the canonical commutator relations (\ref{c0}) and (\ref{c1}). 

One of the first attempts in this direction have been proposed by Segal \cite{S60} and were later developed by \'{S}niatycki \cite{S80}, Doebner, Tolar and Nattermann \cite{D96}\cite{D01}. Without to go in any detail (a review can be found in \cite{A04}) a generalized momentum operator is obtained by projection on a given smooth vector field $X$ on $M$ according to 
\begin{eqnarray}\label{px}
	P_{\!_X} = -i\hbar \left(\nabla_X+\frac{1}{2}\,\text{div} X\right),
\end{eqnarray} 
where $\text{div} X$ is the covariant divergence of the vector field $X$. A straightforward computation (that is here omitted) yields the general commutation relation,
\begin{eqnarray}\label{gencom}
	\left[ P_{\!_X},P_{\!_Y} \right] = -i\hbar\, P_{\!_{\left[X,Y\right]}}
\end{eqnarray} 
where $\left[X,Y\right]$ denotes the commutator of the two vector fields $X$ and $Y$ in the usual sense of the theory of manifolds \cite{S60}. In the case of a linear manifold this vanishes for two infinitesimal translations, and (\ref{gencom}) specializes to the commutativity of the conventional
linear momenta. 

Apparently, operator (\ref{pdw}) of De-Witt can be recovered for vector fields $X$ of the special form $X_k=\partial_k$, with $\Gamma^j_{jk}=\partial_k \log\sqrt{g}\equiv\Gamma_k$. However, that means $\text{div}X_k=\Gamma_k\neq 0$ such that the vector fields $X_k$ cannot be Killing vectors if the underlying manifold is intrinsically curved. \\

The paper is organized as follows. In Sec.\,II, the question of hermiticity of momentum operators (\ref{px}) is discussed. This draws attention to the special importance of Killing frame fields. A classification of manifolds with such a structure and the corresponding Lie algebra is discussed in Sec.\,III. Possible examples $S^1, S^3$ and $S^7$ are considered in Sec.\,IV. In Sec.\,V., a covariant position operator on $S^3$ is defined in terms of geodesic normal coordinates and the associated commutator relations of position and momentum are established. Finally, a summary and outlook is given in Sec.\,VI.

\section{II.\, Hermitean momentum operators}

Let us consider the Hilbert space of square integrable complex functions $L^2(U,\mu)$ on a compact subset, $U\subseteq M$ with smooth boundary $\partial U$ endowed by the inner product,
\begin{eqnarray}\label{}
	(f,g)=\int_U\,d\mu\,f^*g,\quad f,g\in L^2(U,\mu),
\end{eqnarray} 
where $\mu$ is the standard volume measure on $U\subseteq M$.
The statement that momentum operators are Hermitean with respect to an inner product is typically based on the assumption that the boundary terms vanish after partial integration.
Indeed, one cannot define a momentum operator on a bounded domain without specifying boundary conditions. In mathematical terms, choosing the boundary conditions amounts to choosing an appropriate domain for the operator. If one uses no boundary conditions, too many functions are eigenvectors and so the spectrum of $P_{\!_X}$ is the whole complex plane. On the other hand, if Dirichlet boundary conditions are imposed, the situation is too restrictive and one cannot find an orthonormal basis. Thus, if the functions $f\in L^2(U,\mu)$ are smooth on $U$ but constant functions at $\partial U$, in this case finding a domain such that $P_{\!_X}$ is self-adjoint is a compromise to obtain an orthonormal basis of countable spectrum. 
In what follows, the focus will be on the Hilbert spaces ${\cal H}_{U}\subset L^2(M,\mu)$, with 
\begin{eqnarray}\label{dom}
	{\cal H}_{U}= \left\{f\in C^1(U): f|_{\partial U} =\text{const.} \right\} 
\end{eqnarray} 
At this point one has to check whether momentum operator (\ref{px}) still remains Hermitean because the elements in ${\cal H}_{U}$ are not supposed to vanish at the boundary.\\
\\
{\bf Proposition\,1.} Let $X$ be a smooth and divergenceless vector field on $U\subseteq M$. Then $P_{\!_X}$ is Hermitean on ${\cal H}_U$.\\
\\
{\bf Proof.} Let $f,h\in {\cal H}_U$. The divergence of the product $f\,\text{X}$ can be written as 
\begin{eqnarray}\label{pr}
	\text{div}(f\text{X}) = f\,\text{div}(\text{X})+\nabla_\text{x} f.
\end{eqnarray}
On the other hand, one has the decomposition
\begin{eqnarray}\label{}	
	h^*\nabla_\text{x} f&=&\langle h^*\text{X},\text{grad} f\rangle\nonumber\\
	&=&\text{div}(fh^*\text{X})-f \text{div}(h^*\text{X}).
\end{eqnarray}
This equation can be integrated with respect to the volume form $d\mu$ on $U$ as follows:
\begin{eqnarray}\label{x1}
	(h,\nabla_\text{x} f)
	&=&-\int_U d\mu\, f \,\text{div}(h^*\text{X})\nonumber\\
	&&+\int_{\partial U} d\mu_{\partial} \,f  h^*\langle \text{\text{X}},\nu\rangle.
\end{eqnarray}
Here, Stokes' theorem is applied, where $d\mu_{\partial}$ is the volume measure with respect to the boundary $\partial U$ and $\nu$ is the non-negative outward normal on $\partial U$. 
Now, since $f$ and $h$ are assumed to be constant at the boundary of $U$, they can be taken out of the integration in (\ref{x1}) and one can apply Stokes' theorem once more such that the remaining boundary integral on the right-hand side in (\ref{x1}) becomes  
\begin{eqnarray}\label{boundary}
	\int_{\partial U}\!\! d\mu_{\partial} \,\langle \text{X},\nu\rangle=\int_U d\mu\,\, \text{div}X.
\end{eqnarray}
With the assumption $\text{div}\, \text{X}=0$, it follows 
\begin{eqnarray}\label{}
	(h,\nabla_\text{x} f)  = -\int_U d\mu\, f \,\text{div}(h^*\text{X}).
\end{eqnarray}
Finally, eq. (\ref{pr}) is applied and one gets  
\begin{eqnarray}\label{divX}
	(h,\nabla_\text{x} f) + (\nabla_\text{x} h, f) =\!-\!\!\int_U\!\! d\mu\, h^*\!f\,\text{div}(\text{X}).
\end{eqnarray}
The term on the right-hand side can be absorbed into each term on the left-hand side with a prefactor 1/2. By multiplication of the equation with $-i\hbar$, and after applying definition (\ref{px}), we finally obtain 
\begin{eqnarray}\label{sym}
	(h,P_\text{x} f) = (P_\text{x} h, f)
\end{eqnarray}
for all $f,h\in {\cal H}_U$.$ \hfill\square$\\

At this point, to emphasise is that the boundary term in (\ref{boundary}) is not necessarily zero under more general conditions. However, the divergence criterion of the vector field $X$ is sufficient to ensure hermiticity under the given conditions. The possible manifolds which are available under these circumstances will be discussed in the next Section. 

\section{III.\, Geodesic momentum operators}

The classification of manifolds which are compatible to the conditions of Proposition 1 can be described by the following definition of Killing frames:\\
\\
{\bf Definition\,1.} (Killing frame) \cite{AN68}\\
A Riemannian manifold $M$ is said to have the {\it Killing property} if, in some neighborhood of each point of $M$, there exists an orthonormal frame, $X_1,...,X_n$ such that each $X_i$, $i= 1,..., n$, is a Killing vector field (local infinitesimal isometry). Such a frame will be called a {\it Killing frame}.\\

Since any linear combination, with constant coefficients, of Killing vector fields is again a Killing vector field, a manifold has the Killing property if and only if it is always possible to find frames consisting of Killing vector fields, such that $\langle X_i,X_j\rangle = const$, for each choice of $i$ and $j$.
The normality condition of the definition implies that the integral curves of the isometries are geodesics, since a necessary and sufficient condition for this is that the Killing vector fields have
constant length (\cite{S54}, p.\,349; \cite{Y57}, p.\,50).  \\

For instance, let $x, y$ and $r, \varphi$ denote Cartesian and polar coordinates on the Euclidean plane $\mathbb{R}^2$ endowed with Euclidean metric. Then the Killing vector fields corresponding to translations and rotations are $X_1=\partial_x$, $X_2=\partial_y$ and $X_3=\partial_\varphi$. Their squared vector norms are $X_1^2=1,X_2^2=1$ and $X_3^2=r^2$. The Killing vector fields $X_1$ and $X_2$ have a constant length on the whole plane. Their trajectories are straight lines, which are geodesics. The Killing trajectories corresponding to rotations $X_3$ are concentric circles around the origin. The length of $X_3$ is constant along the circles, but non-constant on the whole plane. The corresponding Killing trajectories are circles, which are not geodesics \cite{K16}.\\  
\noindent
In the context given so far one comes to the following:\\
\\
{\bf Definition 2.} (Momentum operator)\\
Let $M$ be an $n$-dimensional Riemannian manifold with Killing frame $X_1,...,X_n$ on $M$. The set of operators defined by  
\begin{eqnarray}\label{pxx}
	P_{{ }_{X_k}}=-i\hbar X_k
\end{eqnarray}
$k=1,...,n$, are called (geodesic) momentum operators in the direction $X_k$ on $M$. \\

This definition is compatible with (\ref{px}) since every Killing vector field $X_i$ is a priori divergenceless, i.e., $\text{div} X_i=0$, $i=1,...,n$. Moreover, the Lie bracket of two Killing fields is still a Killing field. The momentum operators (\ref{pxx}) thus form a Lie subalgebra of vector fields on $M$. If $M$ is a complete manifold, this is the Lie algebra of the translation group. 
In this case the commutation relations of the Killing vector fields are given by  
\begin{eqnarray}\label{com}
	[\,X_i, X_j] =c^k_{ij} X_k 
\end{eqnarray}
where the structural coefficients $c^k_{ij}$ express the multiplication of pairs of vectors as a linear combination. The corresponding commutator relations of the momenta are obtained by multiplication with the physical units $(-i\hbar)^2$ on both sides of (\ref{com}) and subsequently applying the definition (\ref{pxx}), i.e. 
\begin{eqnarray}\label{pcom}
	[\,P_{{ }_{X_i}}, P_{{ }_{X_j}}] =-i\hbar\, c^k_{ij} P_{{ }_{X_k}} 
\end{eqnarray}
which is compatible with the general expression (\ref{gencom}).
The associated Casimir element of this Lie algebra is given by \cite{AN68}: \\
\\
{\bf Proposition 2.} Let $M$ be an $n$-dimensional manifold and $X_1,...,X_n$ be a Killing frame on $M$. There is a decomposition of the Laplace-Beltrami operator, such that    
\begin{eqnarray}\label{p3}
	\sum_{j=1}^n P^2_{\!_{X_j}} = - \hbar^2\Delta.    
\end{eqnarray} 
{\bf Proof.} The vector fields, $X_j$, can be expressed as a linear combination of the coordinate vector fields, $\partial_\alpha=\partial/\partial x^\alpha$, with the Greek letters denoting indices of the local chart, such that
\begin{eqnarray}\label{xi}
	X_i=\xi_i^\alpha\partial_\alpha
\end{eqnarray}
where each $\xi_i^\alpha$ is a function. For every smooth $f$ on $M$, one can write 
\begin{eqnarray}\label{dd}
\sum_{j} X_j^2f	&=&\partial_\alpha(g^{\alpha\beta} \partial_\beta f)- \delta^{ij}(\partial_\alpha\xi^\alpha_i)
\xi^\beta_j \partial_\beta f 
\end{eqnarray}
which has been obtained by the product rule of differentiation. On the other hand, the Laplace-Beltrami operator in the natural frame is
\begin{eqnarray}\label{LB}
\Delta f = \partial_\alpha(g^{\alpha\beta}\partial_\beta f) +\frac{1}{\sqrt{g}} (\sqrt{g}),_\alpha g^{\alpha\beta} \partial_\beta f	
\end{eqnarray}
Now, it follows that expression (\ref{dd}) is equal to (\ref{LB}) for every $f$, if it can be shown that  
\begin{eqnarray}\label{DLBX2}
	\frac{1}{\sqrt{g}} (\sqrt{g}),_\alpha g^{\alpha\beta} +\delta^{ij}(\partial_\alpha\xi^\alpha_i)\,
	\xi^\beta_j =0.	
\end{eqnarray}
Using the basic property, $\nabla g=0$, of the Levi-Civita connection together with $\delta^{ij}\xi_i^\alpha\xi_j^\beta=g^{\alpha\beta}$, one obtains the following condition 
\begin{eqnarray}\label{dxi}
	\delta^{ij}(\nabla_{\!{}_{X_i}}\xi^\alpha_j)=0.
\end{eqnarray}
This identity can be confirmed as follows:
\begin{eqnarray}\label{} 
0	&=&\nabla_{\!{}_{\partial_l}} g^{lk}\nonumber\\
	&=&\delta^{ij}\,\nabla_{\!{}_{\partial_l}}(\xi^l_i\,\xi^k_j)\nonumber\\
	&=&\delta^{ij}\Big(\text{div} X_i\,\, \xi^k_j\,+\nabla_{\!{}_{X_i}}\xi^k_j\,\Big)\nonumber\\
	&=& \delta^{ij}\,\nabla_{\!{}_{X_i}}\xi^k_j
\end{eqnarray}
and thus,
\begin{eqnarray}\label{}
	\sum_{j} X_j^2f	= \Delta f.
\end{eqnarray}
With definition (\ref{pxx}), in physical units, one obtains statement (\ref{p3}). $\text{}\hfill\square$\\

According to this decomposition of the covariant Laplacian there is no ambiguity concerning operator orderings of the kinetic energy term. It is also important to know that the decomposition (\ref{p3}) is independent of the particular choice of the orthonormal basis \cite{H15}. Moreover, the commutator of $\Delta$ with the elements $X_j$ of the Lie algebra (\ref{com}) is given by
\begin{eqnarray}\label{}
	[X_j,\Delta]=0. 
\end{eqnarray}

From the mathematical point of view, the Casimir element has a meaning only for the theory of representations, but not as an element of the Lie algebra, since the product in (\ref{p3}) is not defined for
the algebra itself. However, from linear algebra we know that the eigenvectors of a linear operator always form a basis for the vector space in question. In addition, for any Lie group, one or more of the generators can be simultaneously diagonalized using similarity transformations. The set of generators that can be diagonalized simultaneously are called Cartan generators. Thus, a suggestive and particularly easy basis for the vector space of each representation is given by the eigenvectors of the Cartan generators (see below). \\ 

The scope of the concept given so far asks for a mathematical classification of manifolds with Killing property. It has been shown that a Riemannian manifold having the Killing property must be locally symmetric \cite{AN68}. Thus, each point of a connected Riemannian manifold having the Killing property has an open neighbourhood which is isometric to an open neighbourhood in a simply connected Riemannian symmetric space $M$.
Then $M$ also has the Killing property and, moreover, has global Killing frames. 
In fact, a local Killing frame exists on $M$ because of the given local isometry, and can be extended uniquely to give a global Killing frame. The extension of each Killing vector field to a global Killing vector field is possible since the symmetry implies completeness. This extension remains orthonormal since the Riemannian structure on $M$ is subordinate to a real analytic Riemannian structure (cf. \cite{W67} p.\,240, \cite{H62} p.\,187). 
A simply connected Riemannian symmetric space is said to be irreducible if it is not the product of two or more Riemannian symmetric spaces. It can then be shown that any simply connected Riemannian symmetric space is a Riemannian product of irreducible ones. 

Therefore, the calculations furthermore are restricted to the irreducible, simply connected Riemannian symmetric spaces. Any simply connected Riemannian symmetric space $M$ is of one of the following three types: ($i$) Euclidean type: $M$ has vanishing curvature, and is therefore isometric to a Euclidean space. ($ii$) Compact type: $M$ has nonnegative (but not identically zero) sectional curvature. ($iii$) Non-compact type: $M$ has non-positive (but not identically zero) sectional curvature. Actually, strictly negatively curved manifolds imply that there are no nontrivial (real valued) orthonormal Killing fields.

Manifolds of constant positive curvature are known \cite{AN68} to have the Killing property only if the dimension of $M$ is equal to 1, 3 or 7. For the spheres $S^1$, $S^3$ and $S^7$, in fact, there is a global Killing frame. The construction depends essentially on the existence of a multiplication in $\mathbb{R}^2$ (complex numbers), $\mathbb{R}^4$ (quaternions), and $\mathbb{R}^8$ (Cayley numbers). 

\section{IV.\, Applications} 

From the discussion of the previous Section and the particular role of Killing frames it is straightforward to consider the designated cases of constant curvature manifolds $S^1$, $S^3$ and $S^7$ in more detail. Let us begin with the "trivial" case $S^1$. \\
\\
\noindent{\bf Circle.} 
For the circle one can take $X_1$ to be the unit tangent vector field, say pointing in the anti-clockwise direction. More precisely consider the situation of a compact subset $M\subset S^1$ embedded in $\mathbb{R}^2$. The general solution of the Killing equation ${\cal L}_{X_1}g=0$ on $S^1$, with metric $ds^2=\rho^2 d\varphi^2$ is given by $X_1=\xi^\varphi\, \partial_\varphi$, for $\xi^\varphi\in\mathbb{R}$. Let $\xi^\varphi=1/\rho$, where $\rho$ is the constant (hyper-) radius of the circle, then we have $X_1^2=1$ and the Killing trajectory is a geodesic. The associated momentum operator is 
\begin{eqnarray}\label{d1}
	P_\varphi =-i\hbar\,\frac{1}{\rho} \frac{\partial}{\partial\varphi} .
\end{eqnarray}
This operator is symmetric on any compact set $M\subset S^1$, with $f=\text{const.}$  on the boundary. 
In \cite{GP04}, it is reported that, in quantum mechanics on a circle with standard commutation relation for $\varphi$ and $p_\varphi$, the uncertainty relation cannot be stronger than $\sigma_p\sigma_\varphi\geq 0$, where $\sigma_\varphi$ and $\sigma_p$ are the standard deviations of position and momentum. Indeed, this inequality is not informative at all, since a product of two nonnegative values cannot be negative. Alternatively, one is referred to the approach in \cite{TS09}, which is not affected by difficulties arising in defining a proper measure of position uncertainty on manifolds mentioned in \cite{T04}. By applying the substitution $r=\rho\,\varphi$, which corresponds to the arc-length on $S^1$, the uncertainty principle of \cite{TS09} is given by 
\begin{eqnarray}\label{Un4}
	\sigma_p \Delta r\geq \pi\hbar
\end{eqnarray}
where $\Delta r$ is the measure (length) of a compact domain on $S^1$.  

Before turning over to the case of $S^3$, let us briefly mention that indeed each single component $L_1,L_2,L_3$ of the ordinary textbook angular momentum operator $\mathbf{L}$ is a Killing vector on $S^2$ and moreover, $\mathbf{L}^2=L_1^2+L_2^2+L_3^2$ actually corresponds to the Laplace-Beltrami operator on $S^2$. Although this seems quite promising, these vector fields are not normalizable. Actually, all vector fields on the 2-sphere are inappropriate for this purpose because of the hairy ball theorem of differential topology, which states that there is generally no nonvanishing continuous tangent vector field on even-dimensional $n$-spheres. This makes it hard to think about what kind of vector fields should be appropriate for an adequate description of momentum operators on $S^2$. The discussion in literature about what momentum operators on $S^2$ might be considered to be appropriate extends up to the present day.   \\
\\
\noindent{\bf 3-sphere.} The 3-sphere of radius ${R>0}$ can be understood as the three-dimensional hypersurface in the four-dimensional Euclidean space. This can naturally be described by standard spherical coordinates of $\mathbb{R}^4$, given by \cite{GP09}:
\begin{eqnarray}
	x^1&=&R\cos\chi\nonumber\label{hs1}\\
	x^2&=&R\sin\chi \cos\theta\label{hs2}\nonumber\\
	x^3&=&R\sin\chi \sin\theta \cos\varphi\label{hs3}\nonumber\\
	x^4&=&R\sin\chi \sin\theta \sin\varphi.\label{hs4}\nonumber
\end{eqnarray}
In order to cover all points of the 3-sphere with both positive and negative values of the coordinates $x^i$, it is necessary that $0\leq\chi,\theta\leq\pi$, $0\leq\varphi< 2\pi$. In these coordinates the metric of $S^3$ takes the form 
\begin{eqnarray}\label{metric}
ds^2&=&R^2 \big(d\chi^2+\sin^2\!\chi\,(d\theta^2+\sin^2\!\theta\,d\varphi^2)\big).
\end{eqnarray}
The corresponding Killing equation is solved for the unit sphere ($R=1$) and the following orthonormal Killing frame is selected:
\begin{eqnarray}
	X_1 &=&\phantom{,}\sin\theta \cos\varphi\, \partial_\chi\label{sol1} \\
		&+&(\cot\chi \cos\theta \cos\varphi -\sin\varphi )\, \partial_\theta\nonumber\\ 
		&-&(\cot\chi \csc\theta\sin\varphi+\cot\theta \cos\varphi )\, \partial_\varphi\nonumber\\
	\nonumber\\
	X_2 &=&\phantom{,}\sin\theta \sin\varphi\, \partial_\chi \label{sol2}\\
	&+&(\cot\chi \cos\theta \sin\varphi+\cos\varphi )\, \partial_\theta\nonumber\\ 
	&+&(\cot\chi \csc\theta \cos\varphi -\cot\theta \sin\varphi)\, \partial_\varphi\nonumber\\
	\nonumber\\
	X_3 &=&\cos\theta\,\partial_\chi -\cot\chi \sin\theta\,\partial_\theta + \partial_\varphi\label{sol3}
\end{eqnarray}
Case $R\neq 1$ can be obtained by division of the right side by $R$. The orthonormality relation $g(X_i,X_j)=\delta_{ij}$ is easily verified. The corresponding representation in Cartesian coordinates, $p=(x^1,x^2,x^3,x^4)$, of the Euclidean embedding space $\mathbb{R}^4$ is also determined and given by 
\begin{eqnarray}\label{}
	X_1(p)&=&(-x^4,-x^3,\phantom{-} x^2,\phantom{-}x^1)\\
	X_2(p)&=&(\phantom{-}x^3,-x^4,-x^1,\phantom{-}x^2) \\
	X_3(p)&=&(-x^2,\phantom{-} x^1,-x^4,\phantom{-}x^3)
\end{eqnarray}
which satisfy $X_i(p)\cdot X_j(p)=\delta_{ij}$ with respect to the Euclidean scalar product. One can also check that the Lie algebra generated by $\{X_1,X_2,X_3\}$ is given by the commutation relations, 
\begin{eqnarray}\label{Xij}
[\,X_i, X_j] =-\frac{2}{R}\, \epsilon_{ijk} X_k 
\end{eqnarray}
where $\epsilon_{ijk}$ is the Levi-Civita symbol in three dimensions. 
In physical units this can be rewritten as 
\begin{eqnarray}\label{LieAlg}
	[\,P_{\!{ }_{X_i}}, P_{\!{ }_{X_j}}] =\frac{2i\hbar}{R}\,\,\epsilon_{ijk} P_{\!{ }_{X_k}} 
\end{eqnarray}
The corresponding Hamilton operator, $H$, of a free particle is given by  
\begin{eqnarray}\label{H0}
	H= \frac{1}{2 m}\sum_{i=1}^3 P_{\!{ }_{X_i}}^2 = -\frac{\hbar^2}{2 m}	\Delta
\end{eqnarray}
which is equal to the Casimir element of Proposition 2 in three dimensions. Thus, it follows that 
\begin{eqnarray}\label{HP}
	[\,H, P_{\!{}_{X_i}}] =0 
\end{eqnarray}
for $i=1,2,3$. 

An alternative decompositions of of the Laplacian in (\ref{H0}), by using six (non-orthonormal) Killing vector fields instead of three, has been proposed in Santander etal. \cite{SRC12}. One essential point of the approach in \cite{SRC12} is that the structural coefficients of the associated commutator relations are not constants, such that the Hamiltonian cannot be considered as a Casimir element of the operator algebra. This makes the analysis of the eigenvalues and the corresponding eigenspaces quite complicated. Actually, there seems to be no reason why one should regard a decomposition of the free Hamiltonian in terms of additional angular momenta whose integral curves are not geodesics.

Another interesting approach is the momentum-space quantization of a particle moving on the $SU(2)$ group manifold by Guerrero etal. \cite{G20}. Their algorithm also exhibits a proper and unambiguous realization of the basic operators and of the Hamiltonian, which also turns out to be the Laplace-Beltrami operator on $S^3$. Although, the right-invariant generators (62) in \cite{G20} are different from the Killing frame fields $X_i$ introduced above, they are compatible with the algebra given in (\ref{Xij}). However, the question whether the generators in \cite{G20} also form a Killing frame has not been explicitly discussed.\\

The eigenvalues of $H$ can be obtained by the hyperspherical harmonics on $S^3$, which have been discussed as part of investigations of a variety of gravitational physics problems in spaces with the topology of the 3-sphere \cite{L05}\cite{L17}.
According to \cite{L17}, these hyperspherical harmonics on $S^3$ are denoted by $Y^{nlm}$. The integers $n,l$ and $m$ with $n\geq l\geq 0$ and $l\geq m\geq -l$ indicate the order of the harmonic. These harmonics are eigenfunctions of the covariant Laplacian according to 
\begin{eqnarray}\label{L0}
	\Delta Y^{nlm} = -\frac{n(n+2)}{R^2}\,  Y^{nlm}.
\end{eqnarray}
The corresponding energy eigenvalues $E_n$ of $H$ are given by 
\begin{eqnarray}\label{E0}
	E_n= \frac{\hbar^2}{2 m}\frac{n(n+2)}{R^2}.
\end{eqnarray}
for $n=0,1,...$ 

Now let us consider the corresponding eigenvalue spectrum of the momentum operators. Although $P_{\!{}_{X_i}}$ and $H$ are commuting Hermitian operators it is not necessarily given that each eigenbasis of $H$ is also an eigenbasis of $P_{\!{}_{X_i}}$. Indeed, most of the functions $Y^{nlm}$ given in (\ref{L0}) are not eigenfunctions of the momentum operator $P_{\!{}_{X_3}}$. A simultaneous eigenbasis can be obtained by applying the standard textbook formalism, but based on the specific algebra given by (\ref{LieAlg}). Rather then working with the operators $P_{\!{}_{X_1}}$ and $P_{\!{}_{X_2}}$, it is convenient to work with the non-Hermitian linear combinations, 
\begin{eqnarray}\label{pm1}
	P_{\!{}_{\pm}}=P_{\!{}_{X_1}}\!\! \pm i P_{\!{}_{X_2}}   
\end{eqnarray}
where by definition $(P_{\!{}_{-}})^\dagger=P_{\!{}_{+}}$. Using (\ref{LieAlg}) and (\ref{HP}), it is straight forward to show that 
\begin{eqnarray}\label{}
	\left[P_{\!{}_{+}}, P_{\!{}_{-}}\right]&=& 4\, \frac{\hbar}{R}\,P_{\!{ }_{X_3}}\\
	\big[ P_{\!{ }_{X_3}}, P_{\!{}_{\pm}}^k \big] &=&\pm\,2\, k\,\frac{\hbar}{R}\,P_{\!{}_{\pm}}^k  \label{Pn}
\end{eqnarray}
for $k=0,1,2,...$. Certainly, one also has
\begin{eqnarray}\label{LieAlg2}
	\big[ H, P_{\!{}_{\pm}}^k \big] =0. 
\end{eqnarray}
In order to obtain a simultaneous eigenbasis of $H$ and $P_{\!{}_{X_3}}$ for every fixed $n\in\mathbb{N}$, let us consider the set of orthogonal states $\{\psi^{nk}_{\!{}_{\pm}}\}_{k=0}^n$ given by applying the "ladder" operators $P_{\!{}_{\pm}}$  according to
\begin{eqnarray}\label{}
\psi^{nk}_{\!{}_{\pm}}=P_{\!{}_{\pm}}^k\, Y^{nn(\mp n)}
\end{eqnarray}
where $k=0,1,...,n$. From this definition it follows that $P_{\!{}_{\mp}}\psi^{n0}_{\!{}_{\pm}}=0$. 
By applying the general commutator rule (\ref{Pn}), it can be seen that all of these states are eigenstates of the momentum operator such that
\begin{eqnarray}
	 P_{\!{ }_{X_3}} \psi^{nk}_{\!{}_{\pm}}&=& \pm\,p_{nk}\,  \psi^{nk}_{\!{}_{\pm}}\\
	 p_{nk}&=&(k-\frac{n}{2}) \,\frac{2\hbar}{R} \label{mom1}
\end{eqnarray}
for $k=0,1,...,n$. The physical interpretation becomes straightforward by recalling that the diameter of the 3-sphere is, by definition, the maximal possible geodesic distance ($\pi R$) between two points on $S^3$. If one applies the original definition of Planck's constant $\hbar=h/2\pi$, the maximal resolution $\Delta p_{nk}=p_{nk+1}-p_{nk}$ of the possible momenta in (\ref{mom1}) is given by 
\begin{eqnarray}
	\Delta p_{nk}=\frac{h}{\pi R} \label{momenta}
\end{eqnarray}
Actually this unit of momentum is corresponding to the de-Broglie wavelength, 
\begin{eqnarray}
\lambda_R=\pi R 
\end{eqnarray}
which is identical to the diameter of the manifold. It is hard to think of higher resolutions than this.

On the other hand, the energy eigenvalues can be verified by applying the commutator (\ref{LieAlg2}), such that we obtain the eigenvalue equations
\begin{eqnarray}\label{}
	H \psi^{nk}_{\!{}_{\pm}}= E_n\psi^{nk}_{\!{}_{\pm}}
\end{eqnarray}
where $E_n$ is given in (\ref{E0}). For numerical purposes it is helpful to know the explicit form of the initial functions $\psi^{n0}_{\!{}_{\pm}}$, which are given by 
\begin{eqnarray}\label{}
\psi^{n0}_{\!{}_{\pm}} = C_{n} \,\sin^n\!\!\chi\, \sin^n\!\theta \,\,\,e^{\mp i n \varphi },
\end{eqnarray}
and the normalization constant is 
\begin{eqnarray}\label{}
C_n = \frac{\sqrt{2^{2 n-1} (n+1)}}{\pi}  \,\frac{n!\, (2 n-1)!!}{(2 n)!}
\end{eqnarray}

The representation of the operators $P_{\!{ }_{X_1}}$ and $P_{\!{ }_{X_2}}$ can be obtained by inverting relation (\ref{pm1}) in order to express them in terms of the ladder operators. This completes the brief analysis of momentum operators on $S^3$.  \\
\\
\noindent{\bf 7-sphere.}  Physical applications involving higher dimensional spheres can be found almost exclusively in the context of $N = 1$ supergravity in $11$ dimensions, which is beyond the scope of this study. 
Therefore, let us notify (for information only) some aspects regarding the approach given so far. The sphere $S^7$, considered as Riemannian manifold embedded in $\mathbb{R}^8$ in the usual way, is also designated to have the Killing property. Explicitly, writing points $p$ in $\mathbb{R}^8$ as column vectors and identifying the tangent spaces to $S^7$ with hyperplanes, the vector fields $X_i(p)$, for $i=1,...,7$, are expressed in the Table\,I below \cite{AN68}. Since $p\cdot X_i(p)=0$ and $X_i(p)\cdot X_j(p)=\delta_{ij}$, this gives a global orthonormal frame on $S^7$, which is also a Killing frame. If $\mathbb{R}^4$ is embedded in $\mathbb{R}^8$ as a subset $x^5=x^6=x^7=x^8=0$, the restrictions of $X_1,X_2,X_3$ yield a Killing frame on $S^3$ corresponding to the previous Section (up to sign conventions). 
\begin{table}[h!]
	\begin{center}
		\begin{tabular}{l|c|c|c|c|c|c|c} 
			{$p$} & {$X_1(p)$} & {$X_2(p)$} & {$X_3(p)$}& {$X_4(p)$} & {$X_5(p)$} & {$X_6(p)$}& {$X_7(p)$}\\
			\hline
			$x^1$& $\phantom{-}x^2$ &$\phantom{-}x^3$ & $\phantom{-}x^4$& $\phantom{-}x^5$& $\phantom{-}x^6$& $\phantom{-}x^7$& $\phantom{-}x^8$ \\
			$x^2$& $-x^1$ &$-x^4$ & $\phantom{-}x^3$& $-x^6$& $\phantom{-}x^5$& $-x^8$& $\phantom{-}x^7$ \\
			$x^3$& $\phantom{-}x^4$ &$-x^1$ & $-x^2$& $-x^7$& $\phantom{-}x^8$& $\phantom{-}x^5$& $-x^6$ \\
			$x^4$& $-x^3$ &$\phantom{-}x^2$ & $-x^1$& $\phantom{-}x^8$& $\phantom{-}x^7$& $-x^6$& $-x^5$ \\						
			$x^5$& $\phantom{-}x^6$ &$\phantom{-}x^7$ & $-x^8$& $-x^1$& $-x^2$& $-x^3$& $\phantom{-}x^4$ \\
			$x^6$& $-x^5$ &$-x^8$ & $-x^7$& $\phantom{-}x^2$& $-x^1$& $\phantom{-}x^4$& $\phantom{-}x^3$ \\
			$x^7$& $\phantom{-}x^8$ &$-x^5$ & $\phantom{-}x^6$& $\phantom{-}x^3$& $-x^4$& $-x^1$& $-x^2$ \\
			$x^8$& $-x^7$ &$\phantom{-}x^6$ & $\phantom{-}x^5$& $-x^4$& $-x^3$& $\phantom{-}x^2$& $-x^1$ \\
		\end{tabular}
		\caption{Orthonormal Killing vector representation at point $p$ on $S^7$ embedded in $\mathbb{R}^8$.}
	\end{center}	\label{tab1}
\end{table}

According to this representation the Killing vector fields $X_i$ of the frame bundle and their associated momentum operators $P_{{\!}_{X_i}}\!\!$ on $S^7$ can be easily expressed in terms of hyperspherical coordinates. Moreover, the calculation of eigenvectors and eigenvalues of the corresponding Laplace-Beltrami operator on $S^7$ is straightforward and can be obtained by projecting harmonic fields in Euclidean $\mathbb{R}^8$ onto the unit sphere.

\section{V.\,Covariant position operator on $\mathbf{S}^3$}
Finally, let us bring up some remarks concerning the notion of position operators. Segal \cite{S60} defines the position operator $Q$ as follows: if $f$ is a general function on $M$, then $Q_{{\!}_{f}}$ is defined as the operation of multiplication by $f$. For real $f,g$, the operators $Q_{{\!}_{f}}$ and $Q_{{\!}_{g}}$ are Hermitian, such that there is no difficulty in verifying the commutation relations
\begin{eqnarray}\label{}
	\left[\,Q_{{ }_{f}},P_{{ }_{X}} \,\right] &=&i\hbar\, Q_{{ }_{X\!f}}\label{pq1}\\
	\left[\,Q_{{ }_{f}},Q_{{ }_{g}}\,\right]&=&0
\end{eqnarray}
In order to define a covariant position operator on $S^3$ it is obvious to consider the notion of geodesic distance. 
Since no point of $S^3$ is particularly distinguished, without loss of generality the origin in hyperspherical coordinates is chosen by the "North Pole" $p=(R,0,0,0)\in\mathbb{R}^4$. 
Geodesic normal coordinates, $q=(q^1,q^2,q^3)$, around this origin are such that every element of $S^3$ can be reached by the exponential map 
\begin{eqnarray}\label{}
	\exp_p:&& \!\!\!\!\!\!\!\!T_pS^3\longrightarrow S^3\\
	\exp_p(X)&=& p\cos\big(\frac{s}{R}\big)+R\sin\big(\frac{s}{R}\big) \frac{X(p)}{||X(p)||}
\end{eqnarray}
with $X=q^i X_i\in T_pS^3$ and the geodesic distance function,
\begin{eqnarray}\label{}
	s: q\longrightarrow s_q=||q||
\end{eqnarray}
where $||\cdot||$ is the Euclidean norm in $T_pS^3$. Let 
\begin{eqnarray}\label{}
X_q=\hat{q}^k X_k
\end{eqnarray}
be the corresponding tangent vector field in the unit direction $\hat{q}=q/s_q$. Then, applying the smooth transition map from the hyperspherical chart to geodesic coordinates 
\begin{eqnarray}
	q^1&=&R\chi \sin\theta \cos\varphi\label{qq1}\nonumber\\
	q^2&=&R\chi \sin\theta \sin\varphi\label{qq2}\nonumber\\
	q^3&=&R\chi \cos\theta\label{qq3}\nonumber
\end{eqnarray}
it follows by straight forward computation that $X_q$ can be expressed by
\begin{eqnarray}\label{}
	X_q=\frac{\partial}{\partial s}.
\end{eqnarray}
Accordingly, a geodesic position operator on $S^3$ is defined by $Q_s$ and the commutator (\ref{pq1}) can be expressed by 
\begin{eqnarray}\label{}
	[\,Q_s,P_{{\!}_{X_{{\!}_q}}}]=i\hbar.
\end{eqnarray}
 
A covariant uncertainty relation of position and momentum that is compatible with this approach and was applied to $S^3$ can be found in \cite{TS18}\cite{TS20}. A further generalization to the case of $S^7$ is also possible, but left to further considerations.
 
\section{VI. \, Summary and Outlook }
The description of momentum operators by Killing vector fields is a long-established concept in momentum-space quantization on differentiable manifolds. On the other hand, from classical general relativity, the fundamental importance of geodesic trajectories as a key concept of the theory is also known.

If one wishes to unify these two concepts together in the approach of momentum-space quantization, this leads to the notion of Killing frames on manifolds. These special frames are implicitly part of the classical Cartan formalism. However, the orthonormal frame fields in the Cartan approach are usually not provided as Killing vector fields.
A fundamental principle in the Cartanian approach is to chose the moving frames most suitable to the particular problem. 
Some consequences for the possible manifolds arising from the additional Killing frame property have been discussed. 

The present study is mainly focused on irreducible, simply connected Riemannian manifolds. However, many product manifolds can be constructed out of these irreducible components which also possess the Killing frame property. A straightforward example would be the temporally infinite but spatially finite case $\mathbb{R}\times S^3$. A further generalization is the case of 11-dimensional supergravity with $\mathbb{R}\times S^3\times S^7$. Such possibilities and the extended analysis of the associated spin connections are the subject of further investigations.

\end{document}